# Real-time cardiac cine MRI – A comparison of a diffusion probabilistic model with alternative state-of-the-art image reconstruction techniques for undersampled spiral acquisitions

Authors

Schad[1], Heidenreich[1], Petri[3], Kleineisel[1], Sauer[1], Bley[1], Nordbeck[3], Petritsch[1], Wech[1,2]

[1]Department of Diagnostic and Interventional Radiology, University Hospital Würzburg, Würzburg, Germany
[2]Comprehensive Heart Failure Center Würzburg, Würzburg, Germany
[3]Department of Internal Medicine I, University Hospital Würzburg, Würzburg, Germany

## Correspondence

Oliver Schad
Department of Diagnostic and Interventional Radiology
University Hospital Würzburg
Oberdürrbacher Str. 6, Würzburg, 97080, Germany
E-Mail: schad_o@ukw.de

## Funding Information

This work was supported by the German Ministry for Education and Research under Research Grant 05M20WKA and the Interdisciplinary Center for Clinical Research in Würzburg under Research Grant F-437.

# Abstract

**Purpose:**
ECG-gated cine imaging in breath-hold enables high-quality diagnostics in most patients, arrhythmia and inability to hold breath, however, can severely corrupt outcomes. Real-time cardiac MRI in free-breathing leverages robust and faster investigations regardless of these confounding factors. With the need for sufficient acceleration, adequate reconstruction methods, which transfer data into high quality images, are required.

**Methods:**
Undersampled spiral real-time acquisitions in free-breathing were conducted in a study with 16 healthy volunteers and 5 patients. Image reconstructions were performed using a novel score-based diffusion model, as well as a variational network and different compressed sensing approaches. The techniques were compared by means of an expert reader study, by calculating scalar metrics and difference images with respect to a segmented reference, and by a Bland-Altman analysis of cardiac functional parameters.

**Results**:
In participants with irregular RR-cycles, spiral real-time acquisitions showed superior image quality with respect to the clinical reference standard. Reconstructions using the diffusion model, the variational network and l1-wavelets offered an overall comparable image quality, however sharpness was slightly increased by the diffusion approach. While slightly larger ejection fractions for the real-time acquisitions were exhibited with a bias of 1.6% for healthy subjects, differences in the data acquisition procedure resulted in uncertainties of 7.4%.

**Conclusion**:
The proposed real-time technique enables free-breathing acquisitions of spatio-temporal images with high-quality, covering the entire heart in less than one minute. Evaluation of the ejection fractions using the segmented reference can be significantly corrupted due to arrhythmias and averaging effects, pushing the need for clinically applicable real-time methods. Prolonged inference times of the diffusion model represent the main obstacle to overcome for clinical translation.

# 1 Introduction

Real-time acquisition instead of segmented k-space sampling is advantageous for cardiovascular MRI in many respects[1]. First and foremost, motion artefacts, especially in patients with arrhythmia or who cannot hold breath, can be greatly reduced by not averaging information across several heartbeats[2]. But also for compliant patients with regular cardiac cycles, the comfort of the procedure can be increased considerably by waiving the need for breath-holds, ECG gating and by shortening overall scan times. The increased efficiency can reduce costs and thus indirectly open access to cardiac MRI for a broader patient population[3]. Moreover, the accuracy of investigating dynamic cardiac motion can be enhanced, if data were not pooled from several RR-cycles, which are never 100 % identical and were potentially gated in a suboptimal manner[4]. In more special applications like "exercise MRI", real-time data sampling might be mandatory to match physical information precisely[5]. If data can also be reconstructed in "real time", i.e. with low-latency, also interventional (cardiovascular) procedures like ablation therapies can be guided by real-time MRI with high accuracy[6]. In recent years especially deep learning-based supported approaches have shown promising results in leveraging rapid reconstructions with low-latency[7].

To provide comparable spatial and temporal resolution as typically realized by segmented cine scans, real-time cardiac MRI (cMRI) requires fast data acquisitions. High performance gradient systems and pulse sequences with short TR (FLASH, bSSFP) as well as non-Cartesian k-space trajectories are advantageous for the needed efficiency[8]. The usage of phased-array coils enables the application of Parallel Imaging / Simultaneous Multi-Slice Imaging and thus to sample below the Nyquist-rate. Additional regularization by sparsity models or more recently by means of data-driven priors have further promoted decent image quality despite significant undersampling rates[8]. Especially in iterative reconstructions of non-Cartesian scans, which repeatedly enforce the consistency with measured data, the accuracy of the expected gradient waveforms during acquisition has shown to be of particular importance, as small deviations can amplify and do not cancel out as for full (Cartesian) scans. To this end, correcting trajectories e.g. with a Gradient System Transfer Function has proven to raise image quality[9,10]. Nevertheless, clinical protocols to assess cardiac function etc. are still often dominated by segmented acquisitions. This might be due to the fact that image quality is - on average - often superior compared to state-of-the-art (SOTA) real-time alternatives. In this paper, we therefore compare established PI-, Compressed Sensing- and Variational Network (VN)-models as well as more recently introduced generative diffusion probabilistic models for the reconstruction of functional cardiac MRI scans based on efficient spiral k-space sampling.

Firstly adapted for the reconstruction of undersampled MR acquisition by Hammernik et al.[11], VN's utilize an unrolled network architecture, which alternates between data consistency blocks and (deep) neural network based denoising and artefact removal blocks. A prior in the data reconstruction is induced by pairing data from artefact afflicted acquisitions with a corresponding ground-truth. First implementations employed basic CNN architectures[12], but later works also explored more advanced neural networks architectures such as U-Nets[13], recurrent neural networks[14] or transformer back-bones[15]. Further developments demonstrated VN's fully trained end-to-end[13] along with the integration of gridding operations in the unrolled blocks for the reconstruction of non-Cartesian measurements, as presented by Kleineisel et al.[16] for undersampled spiral acquisitions.

Alternatively, information regarding the underlying data distribution of artefact-free high-quality images can also serve as a prior in the regularization of MR reconstruction algorithms independently from the imaging operator[17]. To do so, several techniques from the realm of generative modeling such as variational auto-encoders (VAE), Generative Adversarial Networks (GAN) and more recently diffusion probabilistic modeling have been demonstrated. The capability of modeling complex structures using VAE's is often limited by the network complexity and therefore tend to create low-fidelity samples. GAN's have shown to be able to achieve great performance, in terms of modeling complex structures, but commonly exhibit instabilities when trained[18].

Emerging diffusion probabilistic models allow to learn the data distribution prior in a straight forward, adversarial-free manner due to the underlying fundamental mathematical basis. Being able to achieve stable and high-quality image generation, they have also attracted attention for medical imaging, with first applications for accelerated MRI reconstruction showing promising results[19–21]. Typically, the iterative nature of diffusion models results in prolonged sampling times, which can be an issue for short latency applications. However, acceleration techniques are currently investigated. Chung et al.[22] propose shortening of the sampling chain using an intermediate estimate as the starting point of the inference. Zhao et al.[23] and Güngor et al.[24] combine diffusion priors with adversarial training for rapid inference.

While previous works mainly exploited data from openly available databases, which were undersampled retrospectively, here we train a diffusion model using spiral cardiac MR-data from scratch. We first describe a diffusion probabilistic approach[20,25,26] to learn the distribution of the subspace of cine MRI scans with high fidelity and subsequently propose a method to exploit this information to regularize a physics driven reconstruction of real-time acquisitions in the beating heart. In a study with healthy volunteers and patients suffering from arrhythmic cardiac diseases, the method is then compared against alternative techniques like the aforementioned Variational Networks[11,13,16], a low rank plus sparse algorithm[27], as well as total variation- and wavelet-based compressed sensing[28].

## 2 Methods

MRI can be understood as a linear inverse problem $y = Ax$, where the image $x \in \mathbb{C}^{n \times m}$ has to be retrieved from the complex multi-coil data $y \in \mathbb{C}^{\#coils \times n \times m}$. Typically, the "MRI Operator" $A$ can be disentangled into $A = M\mathcal{F}S$, consisting of coil sensitivities $S \in \mathbb{C}^{\#coils \times n \times m}$, the Fourier transform $\mathcal{F}$ and a sampling mask $M$. For acquisitions with multiple receiver coils (i.e. $\#_{coils} > 1$), parallel imaging can be exploited to reconstruct images from acquisitions $M$, which are "undersampled" by a factor $R < \#_{coils}$ (i.e. the Nyquist sampling criterion is violated in k-space). As measurements with multiple coils are, however, correlated, R can typically not nearly reach $\#_{coils}$, without significantly enhancing image noise. In the case of non-Cartesian sampling, $\mathcal{F}$ needs to incorporate some kind of gridding procedure to transfer data onto a Cartesian grid. We will use the term "naïve" to refer to the undersampled, gridded and coil-combined reconstructions without further regularizations.

To push the potential for acceleration, plenty reconstruction techniques have been proposed on top of parallel imaging by assuming prior knowledge in the underlying data distribution and regularizing the inverse equation accordingly[12,17]. This can be described by the following optimization problem

$$\arg \min_x \tfrac{1}{2} \|Ax - y\| + \psi(x), \qquad \text{[I]}$$

where $\psi(x)$ represents a regularization term.

In the following, we first establish our proposed acquisition scheme for accelerated (real-time) cMRI based on spiral bSSFP. Subsequently, a reconstruction technique based on diffusion probabilistic models and the corresponding training is described. Ultimately, alternative modern reconstruction techniques are briefly outlined.

### 2.1 Data acquisition and gradient waveform correction

To provide fast and efficient data sampling for real-time cardiac MRI, we implemented an undersampled spiral bSSFP-sequence, schematically depicted by the pulse sequence diagram in Fig. 1. Spiral trajectories were created using the variable density spiral algorithm by B. Hargreaves[29], additionally optimized in consideration of peripheral nerve stimulations[30]. Triangular rewinders were appended to the spiral read out to null the zeroth gradient moment. Acquisition parameters are depicted in Table 1.

Banding artefacts were avoided in the shimmed region of interest by using RF phase alternation of 180 ° ($\pm \alpha$). Real-time frames consisted of 13 equidistant spiral arms (acceleration factor R~5) resulting in a temporal resolution of ~48 ms per frame.

To simultaneously reconstruct real-time frames as well as segmented cine, data were first acquired during breath-hold. In this case, the rotation schedule of the spiral arms was individually adjusted for each subject's heart rate. The pattern of 13 equidistant arms (angle increment ~28°) was repeatedly acquired for slightly longer than one heartbeat, as estimated from the "live" ECG of the MR-Scanner. The entire pattern of 13 spiral arms was then rotated by an angle that minimizes the remaining gaps in k-space for all acquisitions accumulated over time.

This sampling strategy was repeated for approximately 9 heartbeats, accounting for one heartbeat in transient phase. Scan times varied between 9 s and 15 s for a single slice. By repetitively sampling the k-space center, a self-gating "ECG" signal can be extracted at k = 0 (i.e. the "DC-signal"). Binning the differently rotated, undersampled real-time frames from matching timestamps in the cardiac cycle, segmented fully sampled cardiac phases can be derived from the spiral acquisition in breath-hold, besides matching real-time frames. Using segmented data from 8 heart beats results in cardiac phases consisting of 104 spiral arms with equidistant angular distribution.

These acquisitions were used for training the DL-based approaches: For the diffusion model, an unsupervised fashion was applied to learn a data distribution prior using the fully sampled cardiac frames only. Contrarily, the VN was trained in a supervised manner, by matching the undersampled real-time frames with the respective fully sampled frame acquired for the same cardiac phase across different RR-cycles. This can be repeated for each of the undersampled real-time frames of the different RR-cycles, which is additionally taking into account varying aliasing artefacts due to the different rotation angles of the k-space trajectory.

Additionally, these measurements were repeated in free-breathing with a shortened scan duration, by rotating the pattern already after a fraction of a RR-cycle. Spiral trajectories were retrospectively corrected using the scanners gradient system transfer function as determined in a different study[10].

All spiral acquisitions were initially transferred onto a Cartesian grid of size 512 px × 512 px using GRAPPA Operator Gridding (GROG)[31,32]. To this end, GRAPPA Kernels were calculated from a fully sampled temporal average across ~8 heartbeats using convolution gridding. Coil sensitivity maps were estimated using ESPIRIT, applied to the temporally averaged k-space using BART[28].

**Table 1**: **Overview of the acquisition parameters of the Cartesian cine reference and the proposed spiral technique.**

| Sequence parameters | Segmented Cartesian cine bssfp | Undersampled spiral real-time bssfp |
|---|---|---|
| Slice thickness [mm] | 8 | 8 |
| Flip angle [°] | 70 | 70 |
| Spatial resolution [mm] | 1.25x1.25 | 1.29x1.29 |
| FOV [mm] | 320x260 | 592x592 |
| Image Matrix [px] | 256x208 | 512x512 |
| TR [ms] | 3.94 | 3.70 |
| TE [ms] | 1.97 | 0.61 |
| Temporal resolution [ms] | 44±4 | 48 |
| Pixel Bandwidth [Hz/pixel] | 528 | 407 |
| GRAPPA | 2 | - |

Note: The actual spatial resolution of the undersampled spiral technique is given by the k-space location with the largest distance to the k-space center during acquisition. Since transferring off-grid data to a 512px x 512px Cartesian grid using GROG introduced slight zero filling, reconstructions displayed pixel sizes of 1.16mm x 1.16mm with corresponding FOV.

## 2.2  Volunteer study

The study was approved by the local ethics committee under licence ID 173/22_skpm, and written informed consent was obtained from each participant. Inclusion criteria were age > 18 years, and for patients a clinical diagnosed intermittent atrial fibrillation. Exclusion criteria were common contraindications to MRI examinations. A total of 16 healthy participants and 5 patients with atrial fibrillation diseases were examined on a 1.5 T clinical whole-body scanner (Siemens Magnetom Avanto[fit]), which provided a maximum gradient amplitude of 40 mT/m and a maximum slew rate of 170 mT/m/ms. Data were acquired in breath-hold and free-breathing. Left-ventricular short-axis slice orientations (SAX) were used, seamlessly covering the entire ventricle from base to apex.

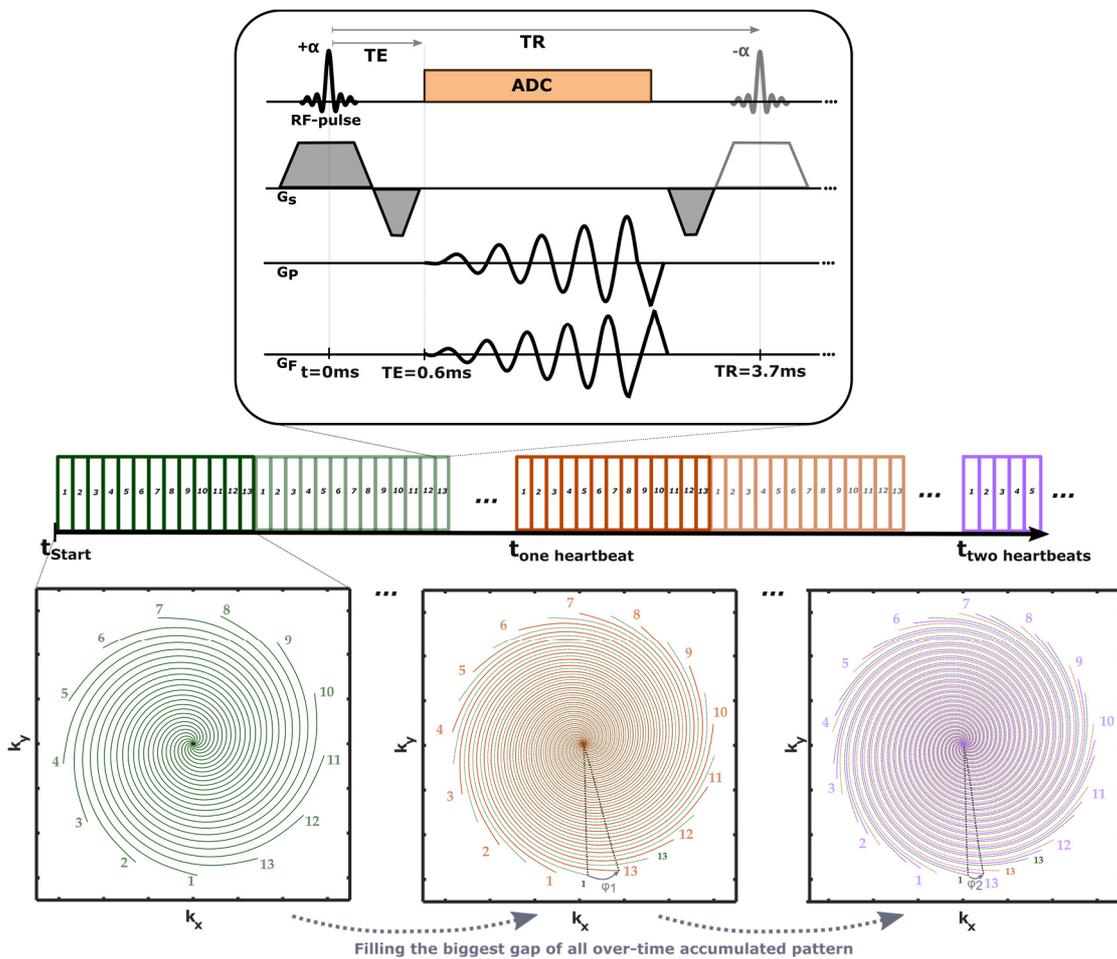

**Figure 1**: Overview of the spiral bSSFP-sequence. Real-time frames contain data from 13 equidistant spiral arms, which were repeatedly acquired for a duration slightly exceeding one RR-cycle in breath-hold (as checked by the "live" ECG of the scanner). After one heartbeat, this pattern was rotated, filling the largest gap in between all previously captured patterns. Sampling for ~9 heartbeats allowed for a segmented reconstruction of fully sampled cardiac phases in addition to real-time acquisitions. For a shortened scan duration in free-breathing acquisitions, the pattern was rotated already after a fraction of an RR-cycle.

In addition to the spiral acquisition scheme described in Sec. 2.1, ECG-gated, breath-hold clinical Cartesian bSSFP acquisitions were performed at identical slice positions as a reference. An overview of the measurement parameters for both methods can be found in Table 1.

## 2.3  Score-based diffusion models

**Vanilla score-based diffusion models**
We follow the description of the generative diffusion process using a score-based approach[25] to model the underlying data distribution of cMRI data. For a more detailed description of diffusion models and their usage as a prior for solving inverse problems see[20,24–26,33,34].
Consider a diffusion process for an image $x(t)$, where $t \in [0,T]$ presents a continuous time variable. $x(0)$ corresponds to a sample of the probability distribution $p_0(x)$ of a given dataset and $x(T)$ is the sample fully perturbed to Gaussian noise. Such process demonstrates a solution to the stochastic differential equation (SDE) of the form: $dx = f(x,t)\,dt + g(t)\,dw$, where w corresponds to Brownian motion.
Reversing the given SDE by starting from Gaussian noise to obtain samples from the data distribution $p_0(x)$ also results in a diffusion process, with the reverse-time SDE (rSDE) following

$$dx = [f(x,t)\,dt - g(t)^2\,\nabla_x \log(p_t(x))]\,dt + g(t)\,dw'. \qquad \text{[II]}$$

Choosing $f(x,t) = 0$ and $g(t) = \sqrt{\frac{d(\sigma(t)^2)}{dt}}$ results in so called variance exploding (VE) SDE, where σ(t) is a positive, monotonically increasing function, typically chosen to be a geometric sequence, such as $\sigma(t) = \sigma_{min}\left(\frac{\sigma_{max}}{\sigma_{min}}\right)^t$ for $t \in (0,1]$.
If the score $\nabla_x \log(p_t(x))$ is defined for all timesteps $t$, one can generate samples from the distribution $p_0(x)$ by solving the rSDE.
Commonly, since the true score is not available, it is estimated by training a time-dependent neural network $s_\theta(x(t), t)$ using denoising score matching[35] and optimizing the following l2-loss function

$$\arg\min_\Theta E_{t,x(0),x(t)|x(0)}[\lambda(t)||s_\Theta(x(t),t) - \nabla_x \log(p_{0t}(x(t)|x(0))||_2^2]. \qquad \text{[III]}$$

Here $\lambda(t)$ corresponds to a positive time-dependent weighting function.
After training the network as an approximation for the score, the rSDE can be solved, e.g. by using numerical methods such as reverse diffusion and Euler-Mayurama sampling, Langevin dynamics or hybrid algorithms such as predictor-corrector methods[36–38], thereby new samples can be drawn from the learned data distribution. To do so, $t$ is typically uniformly discretized into N steps $t_i = i * \Delta N$ with $\Delta N = \frac{1}{N}$ and i ∈ [1, N], which leads to the continuous description in the case for $N \to \infty$. The discretized sampling process of the reverse diffusion sampler to solve eq. II, which we use in this work, is given by $x_{i-1} = x_i - f_i(x_i) + g_i g_i^T s_\Theta(x_i, i) + g_i z$, with z being standard Gaussian noise.
The perturbation with Gaussian noise between two discrete timesteps is defined by the perturbation kernel $p(x_i|x_{i-1}) = N(x_i, x_{i-1}, (\sigma_i^2 - \sigma_{i-1}^2)I)$. Thereby the perturbation of $x_i$ can be described in a single step using the unperturbed input $x_0$ with the kernel $p(x_i|x_0) = N(x_i, x_0, \sigma_i^2 I)$. Using $\nabla_x \log(p_{0t}(x(t)|x(0)) = (x(t) - x(0))/\sigma(t)^2$ and $x(t) = x(0) + \sigma(t) * z$,

the training objective simplifies to $\arg\min_{\Theta} E_{t,x(0),x(t)|x(0)}[\lambda(t)||s_\Theta(x(t),t) - z/\sigma(t))||_2^2]$, where z is standard Gaussian noise.

**Conditioned Sampling**

So far, the description of the diffusion models solely applies for a purely generative task. Solving an inverse problem translates to computing the posterior probability $p(x(t)|y)$ for an image x(t) given the data y, which can be rewritten to $p(x(t)|y) \sim p(x(t))p(y|x(t))$ using Bayes rule. According to Song et al.[25], computing the score corresponds to $\nabla_x \log(p_t(x(t)|y) \approx \nabla_x p(x(t)) + \nabla_x \log(p_t(y(t)|x(t))) = s_\Theta(x(t),t) + \nabla_x \log(p_t(y(t)|x(t)))$. The first term resembles the unconditioned score function, estimated by the trained network. The latter term is typically not tractable, as only the dependence $p(y|x(0))$ is known. In the case of the inverse problem of MRI $p(y|x(0)) = N(y, Ax(0), \sigma_\tau^2 I)$, assuming zero-mean Gaussian noise with standard deviation $\sigma_\tau$ during the measurement. The score then corresponds to $\nabla_x \log(p(y|x(0)) = \frac{A^H(y-Ax)}{\sigma_\tau^2}$. Approximating $p_t(y(t)|x(t))$ is an active research topic and currently no straight forward solution has been found to this problem. In the case of MRI reconstruction different approaches have been proposed. Many of them use the approximation $p_t(y(t)|x(t)) \sim p(y|x(0))$ and yielded positive results by counteracting for the incorrectness using some heuristic weighting in the data consistency term[39]. Recently, more sophisticated methods to solve this task have been proposed, e.g. by Chung et al.[40] and Song et al.[41]. In this work we follow the lines of Jalal et al.[19], aiming for more sophisticated and possibly more optimal approaches in future development of the imaging technique studied here.

**Training for cardiac MR**

In this work, we initially exploited properties of the diffusion model to capture the probability distribution of cardiac MR cine images. Using the weighting function $\lambda(t) = \sigma^2(t)$, the training task was $\arg\min_{\Theta} E_{t,x(0),x(t)|x(0)}[||\sigma(t) s_\Theta(x(t),t) - z)||_2^2]$. The typical geometric series was used for the noise scheduler with $\sigma_{min} = 0.01$ and $\sigma_{max} = 378$. A total of $N = 2000$ diffusion steps were trained using Adam optimizer with a batch size of 1 and a linear learning rate warm up, that reached a constant learning rate of $2x10^{-4}$ after 5000 steps with $\beta_1 = 0.9$ and $\beta_2 = 0.999$ and an exponential moving average rate of 0.999. The time-dependent neural network called noise conditional score network ++ (ncsnpp) proposed by Song et al.[25] was used as score function. This network receives a one channel input image conditioned on a noise scale and outputs an estimate for the overlaying noise in the image.

2120 individual fully sampled (spiral trajectory) 2D frames in magnitude reconstruction of size 512 px × 512 px, which were acquired in 96 slices of 8 healthy participants, were sampled from the study as depicted in Sec. 2.2 and used for training of the diffusion network. Data were normalized to a data range of [0 1]. Training for 100 epochs took roughly 5 days on an RTX A6000 GPU(48Gb). The diffusion model combined with the trained network is capable of generating realistically looking, artificial samples of the initial probability distribution (i.e. cine frames with contrast features etc. compliant with those of the training set, see Fig. 2).

**Artificially generated cine frames from the unconditioned diffusion prior**

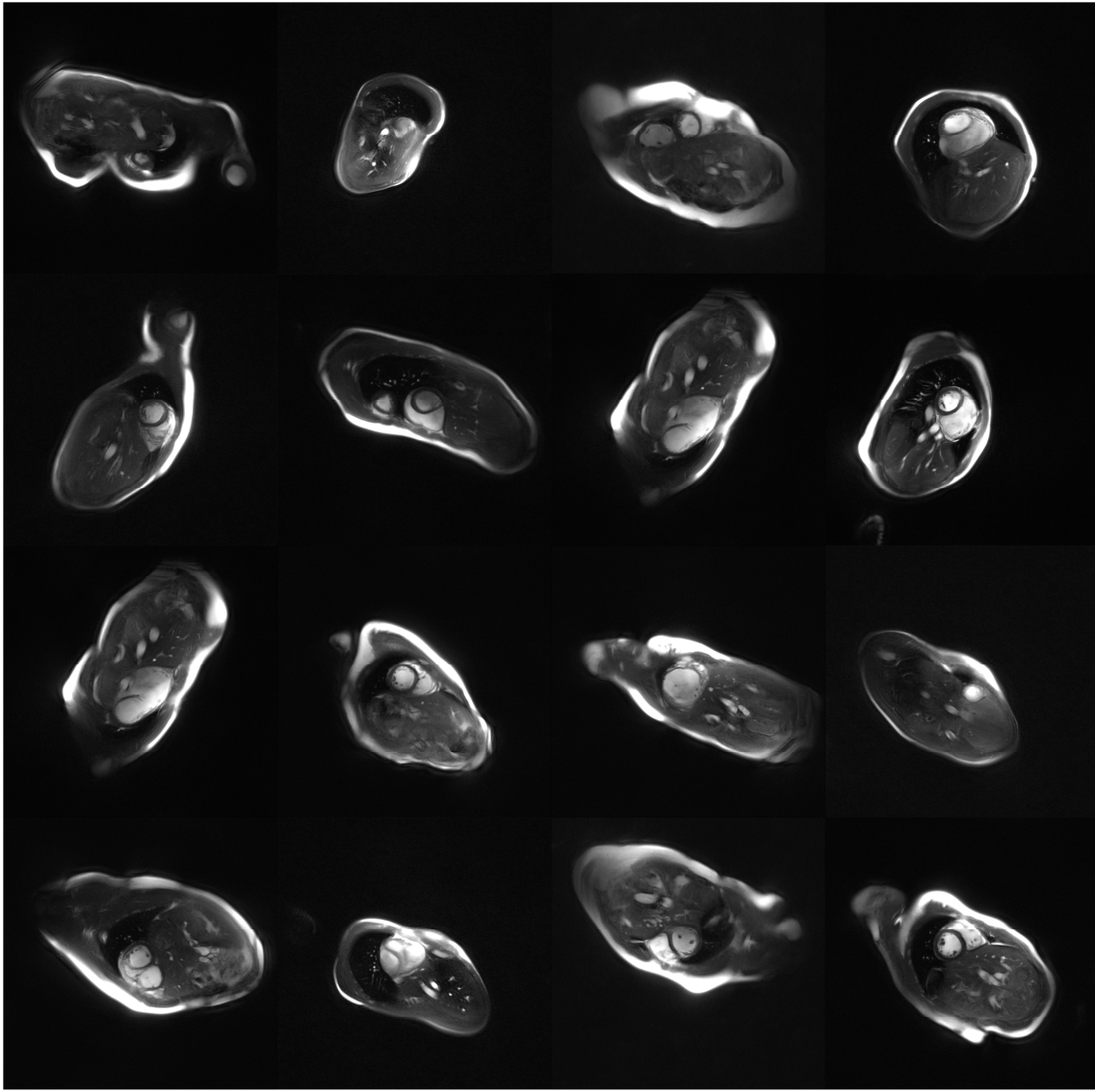

**Figure 2**: Artificial samples from the distribution of cardiac cine frames as provided by the trained diffusion model. Results overall comply with quality, contrast, field-of-view, gross anatomy etc. of the original datasets and its variations. Deviations from realistic anatomy (e.g. first image in last row) do not pose a fundamental problem for the downstream reconstruction scheme, as the additional data consistency term guides the diffusion process towards a result in accordance with the anatomy of the actual examination.

**Conditioned sampling for reconstruction of cardiac MR**
To perform conditioned sampling from undersampled MR data, we employ the reverse diffusion sampler in combination with weighted data consistency terms. A predictor-corrector approach was not employed in favor of sampling speed, since we saw no significant improvement in image quality. The iteration procedure is visualized in Fig. 3 and outlined by Algorithm 1.

Since the network was trained on magnitude data only, a phase estimation that maps magnitude to complex images back and forth during the data consistency step is utilized (instead of separating real and imaginary part of the complex data as in[20]). Additionally, we work with coil-combined images, allowing for faster sampling as opposed to sampling each coil-image individually[20].

This is done by multiplying the image with a phase map $p_{mean}$ estimated from the phase of a fully sampled temporal average image. We added a weighted combination using linearly decreasing scalings $\lambda_i$ with the phase of the image after applying data consistency, such that the phase can dynamically transition into the "true" phase given by the individual real-time frame with short temporal footprint. Subsequently, the data consistency is enforced, comparably to a thresholded Landweber approach. Inspired by Jalal et al.[19], data consistency terms were weighted depending on the discretized iteration step i with $\gamma(i) = \eta/(\sigma^2 + t_i^2)$, with the hyperparameters being set to $\sigma = 0.1$ and $\eta = 0.01$.

In the most straightforward execution, the algorithm would start by drawing a sample from pure Gaussian noise. After application of the unconditioned reverse diffusion sampler, the real-valued data is converted to complex data. This method enables the execution of the data consistency operation with the network being trained on magnitude data only. After transforming the data conditioned output back to real valued data, the procedure is repeated.

Instead of starting the inference process with pure Gaussian noise and running the algorithm for an extended number of time steps, the procedure can be sped up by using an reconstruction estimate perturbed with noise from an earlier diffusion step as outlined in [20,22].

We start with a coil-combined magnitude image as obtained from an iterative SENSE reconstruction normalized to a data range [0 1], additionally forward diffused with noise from step n=100. We ran the diffusion reconstruction for the last n=100 steps, with interleaved discretization steps by setting N=500 during reconstruction, as suggested in [20]. In the final step no additional noise was added in the reverse diffusion sampler.

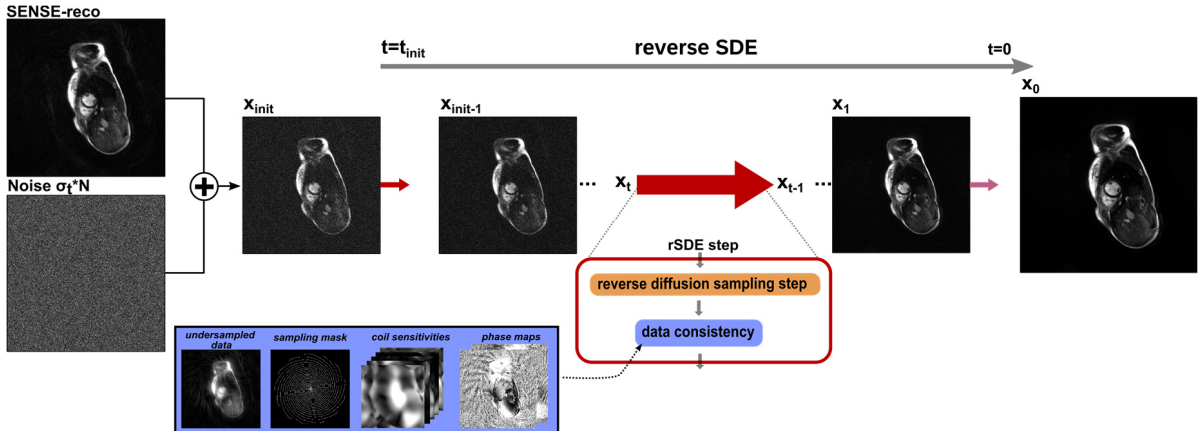

**Figure 3**: Schematic view of the proposed diffusion reconstruction chain for undersampled spiral cMR. Starting from an iterative SENSE reconstruction, the sampling procedure begins by subjecting the magnitude image to a forward diffusion step with Gaussian noise. The reverse SDE is then solved using the reverse diffusion sampler approach[25], which additionally enforces data consistency. For further details see Algorithm 1.

**Algorithm 1: Physics-based image reconstruction exploiting a diffusion probabilistic model**

---

Require: $s_\Theta, n, \{\sigma\}_{i=0}^n, p_{mean}, \{\lambda\}_{i=1}^n, x_{SENSE}, y$

$x_n = x_{SENSE} + \sigma_n \mathcal{N}(0, I)$      # Forward diffusion step

$p_{last} = p_{mean}$

For i=n:1 do:

     $x_{i-1} = x_i + (\sigma_i^2 - \sigma_{i-1}^2) s_\Theta(x_i, \sigma_i)$      # Reverse Diffusion Sampling

     If i!=1: $x_{i-1} = x_{i-1} + \sqrt{\sigma_i^2 - \sigma_{i-1}^2}\, \mathcal{N}(0, I)$

     $x_{i-1} = \lambda_i\, x_{i-1} \odot e^{ip_{mean}} + (1 - \lambda_i) x_{i-1} \odot e^{ip_{last}}$      # Phase mapping

     $x_{i-1} = x_{i-1} + \gamma(i) * A^*(y - Ax_{i-1})$      # Data consistency

     $p_{last} = angle(x_{i-1})$      # Retrieve updated phase image

     $x_{i-1} = real(x_{i-1} \odot e^{-ip_{last}})$      # Invert phase mapping

End for

Return $x_0$

---

Caption: $x_{SENSE}$ corresponds to an iterative SENSE reconstructed magnitude image of the undersampled data, gridded using GROG y. $\sigma_i$ is the noise scale of the i-th diffusion step, in our case starting from the step n=100. Sampling with the trained score function $s_\Theta$ is performed using the reverse diffusion sampler. During data consistency, the final phase of the reconstruction $p_{last}$ is estimated using a linearly weighted combination by the scaling factor $\lambda_i$ with the phase of a temporal average $p_{mean}$.

## 2.4 Variational Networks

Variational Networks (VNs) represent an alternative and already more established data driven solution for the minimization problem in eq. [I] and their performance has already been demonstrated for the reconstruction of undersampled MR[11], and specifically also for cardiac MRI with non-Cartesian trajectories[16]. Eq. [I] is approximated using Landweber-Iterations[42] here. A neural network is used in a cascaded scheme to regularize the reconstruction of undersampled data. Models are trained in a supervised fashion by evaluating a loss function between matched pairs of reconstructions of undersampled acquisitions and fully sampled references.

For the reconstruction of our undersampled spiral cMR data, we used an in-house developed VN[16], which is based on a cascaded U-Net architecture. The unrolled blocks of the VN incorporated gridding operations and coil-(de)combinations, allowing to work on the complex off-grid raw data. Therefore the VN is the only model investigated in this study, which incorporates convolution gridding instead of GROG. Real and imaginary parts of the gridded coil-combined images were normalized to zero-mean and unit standard deviation before feeding them as a two channel input into the network. 20 cascades were used, with a total of 4 pooling layers and 16 channels in the first layer.

Training the VN for 13 epochs using the mean squared error (MSE)-loss took ~6 days on a single RTX A6000 GPU(48Gb). To this end, 16.960 undersampled timeframes from 8 heartbeats from in total 96 slices of 8 healthy volunteers were used together with the reference images from binned cine frames. For validation, 4800 undersampled time frames were paired with respective segmented cine frames of two additional healthy volunteers. A fixed learning rate of $5 \cdot 10^{-3}$ was used for the Adam optimizer. Because the validation loss showed an increase after the 7th epoch, saved network weights from this epoch were used for reconstruction.

## 2.5 Model-based reconstruction techniques – compressed sensing

Compressed sensing models assume sparsity of the data distribution in a specific representation space by integrating this prior knowledge as additional regularizer of the minimization problem. l1-Total Variation (TV) and l1-Wavelet (WV) were applied in this work to reconstruct real-time frames separately (i.e. 2D reconstruction) for further comparison. The openly available BART-toolbox[28] was used for this purpose. Regularization parameters were set to $\lambda = 0.03$ for both approaches (TV and WV). TV employed ADMM optimization with 40 iterations, while FISTA with 30 iterations was used for WV reconstructions. Prior to reconstruction, 2D multi-coil raw-data were normalized with respect to the maximum value of a naïve reconstruction (inverse Fourier transform of the data after applying GROG plus coil combination).

Additionally we used a low rank plus sparse (LRS) model that also enforces sparsity in the temporal domain[27]. Here $N = 60$ iterations were performed using regularization values of $\lambda_L = \lambda_S = 0.02$ for thresholding the singular values of the Casorati matrix (i.e. low-rank approximation) and the temporal frequencies of the residual sparse domain, respectively. Data of the entire time series to be reconstructed were normalized with respect to the maximum value of the respective coil-combined magnitude image series here.

## 2.6 Quantitative evaluation of image quality

Ideally, a quantitative analysis of different reconstruction methods can be performed based on a fully sampled data set that can be retrospectively undersampled. Because it is not feasible to acquire fully sampled "real-time" frames, segmented cine can be used instead. Applying the spiral acquisition scheme as depicted in section 2.1. and in breath-hold, we were able to reconstruct such a fully sampled segmented cine in addition to real-time frames from a single 2D+t series. To retrospectively extract an undersampled k-space, we applied the undersampling mask from a spiral real-time acquisition to the fully sampled cine k-space. Both were gridded using GROG. As the VN was based on convolution gridding rather than GROG, a fully sampled (convolution) gridded reference was compared with an undersampled reconstruction in an analogue manner.

Using measurements from 94 slices of eight healthy volunteers, matching pairs of fully and undersampled frames were generated. Patient data as well as data from two healthy volunteers were disregarded (24 of 94 slices) for this evaluation due to arrhythmia leading to a failure of the data binning procedure and therefore clearly blurred references. Ultimately, a total of 1632 images were utilized. Segmented reference data as well as undersampled reconstructions were cropped to image sizes of 120px × 120px, centered on the myocardium. Individual images in the time-series were rescaled to a data range of [0 1].

We then evaluated the reconstruction methods described in 2.3-2.5 by computing the following metrics: structural similarity index (SSIM), normalized root mean square error (NRMSE) and peak signal-to-noise ratio (PSNR) with respective standard deviations using the skimage.metrics library[43].

## 2.7 Expert reader study

Image quality of free-breathing reconstructions in a temporal series of 40 real-time frames (l1-Wavelet, LRS, VN & diffusion) as well as reference images given by the Cartesian reconstructions were rated by a board certified radiologist (J.F.H.) with seven years of experience in cardiac imaging. To this end, data from 145 SAX oriented slices from 8 healthy volunteers and 5 patients were used. The rating was performed using equidistant five-point rating scales for the following items: Blood/Myocardium contrast (1: excellent, 2: good, 3: moderate, 4: fair, 5: poor), presence of noise which impairs assessment of myocardial structures (1: none, 2: very few, 3: moderate, 4: substantial, 5: non-diagnostic), artefacts that affect the cardiac structures (1: none, 2: very few, 3: moderate, 4: substantial, 5: non-diagnostic), temporal dynamics along the t-dimension (1: excellent, 2: good, 3: moderate, 4: fair, 5: poor) and the sharpness of myocardial contours (1: excellent, 2: good, 3: moderate, 4: fair, 5: poor). Ratings were conducted for the complete slice stack per subject. Ratings were carried out in an encoded and randomized manner.

## 2.8 Quantification of functional parameters

In order to investigate the performance of the proposed real-time acquisition scheme and corresponding reconstruction methods for the evaluation of cardiac function, 145 SAX slices of 5 patients and as well as 8 healthy participant were manually and jointly segmented by an MD candidate under supervision of the aforementioned expert reader as well as the expert reader himself. A dedicated segmentation software for medical images (MEVIS draw, Frauenhofer MEVIS, Bremen, Germany) was used. We compared the Cartesian gold-standard cine acquired in breath-hold to real-time and free-breathing acquisitions, both reconstructed by the diffusion approach. In addition, segmented spiral cine reconstructions from breath-hold acquisitions of 8 healthy subjects were also evaluated for further comparison.

From the segmentations of the endocardial border of the left ventricle the following functional parameters were derived: end-systolic volume (ESV), end-diastolic volume (EDV), stroke volume (SV) and ejection fraction (EF).

Differences in the quantification of functional parameters between ground truth (Cartesian cine), segmented spiral cine and the real-time series were evaluated using Bland-Altman analysis[44,45].

## 2.9 Comparison of reconstruction time

In-house implementations such as the VN, LRS and the diffusion model were implemented and reconstructed on a GPU using the pytorch library[46], whereas the l1-CS models ran on CPU only using BART[28]. Reconstruction times were computed per frame.

# 3 Results

## 3.1 Qualitative comparison of diffusion model with alternative methods

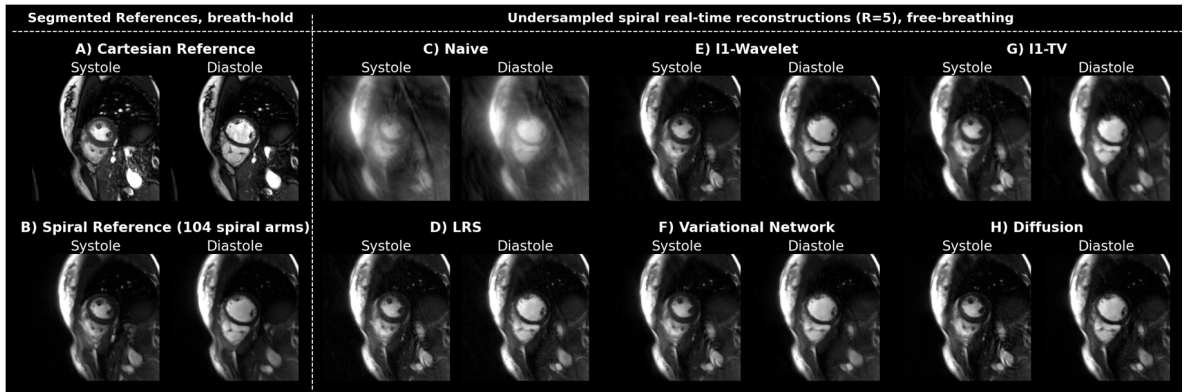

**Figure 4**: Overview of the different acquisition and reconstruction methods for a systolic and diastolic frame of one healthy participant in SAX orientation. A) shows images acquired by a Cartesian bSSFP protocol used in the clinical routine. Here, data are segmented over multiple heart beats to reconstruct cardiac phases for a single (pseudo) heartbeat. A similar averaging method can be applied to the spiral acquisition using "self-gating" (see Sec. 2.1), resulting in the images shown in B). Artefact level of real-time frames reconstructed without additional regularization in the reconstruction is depicted in C). The images D)-H) depict the results for different reconstruction methods applied to real-time acquisitions in free-breathing (temporal footprint 48 ms).

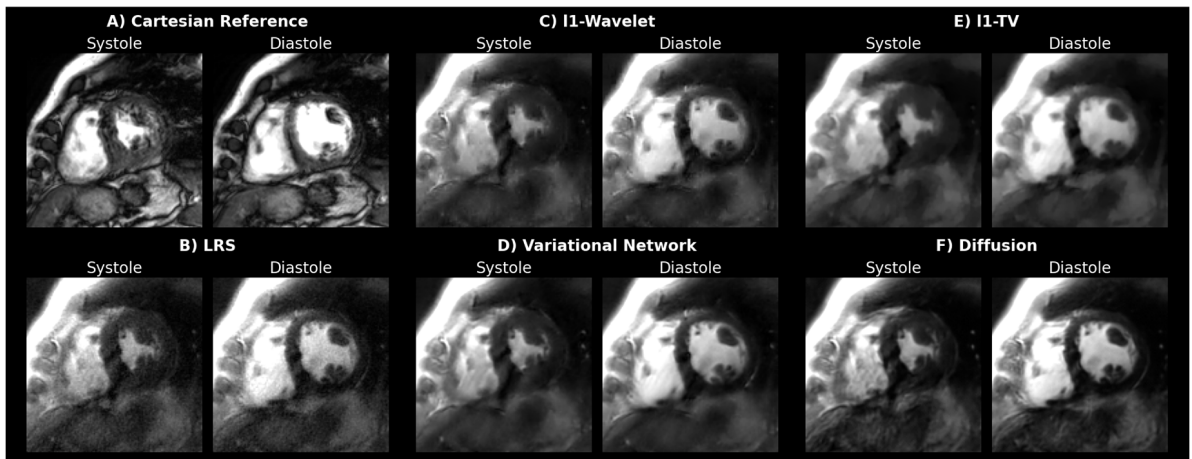

**Figure 5**: Comparison between the Cartesian reference A) and different reconstruction techniques B)-F) applied to spiral acquisitions in real-time free-breathing for a systolic and diastolic frame of one arrhythmic patient in mid-ventricular SAX orientation. Due to varying RR-cycles, the Cartesian method was not capable of resolving cardiac phases accurately.

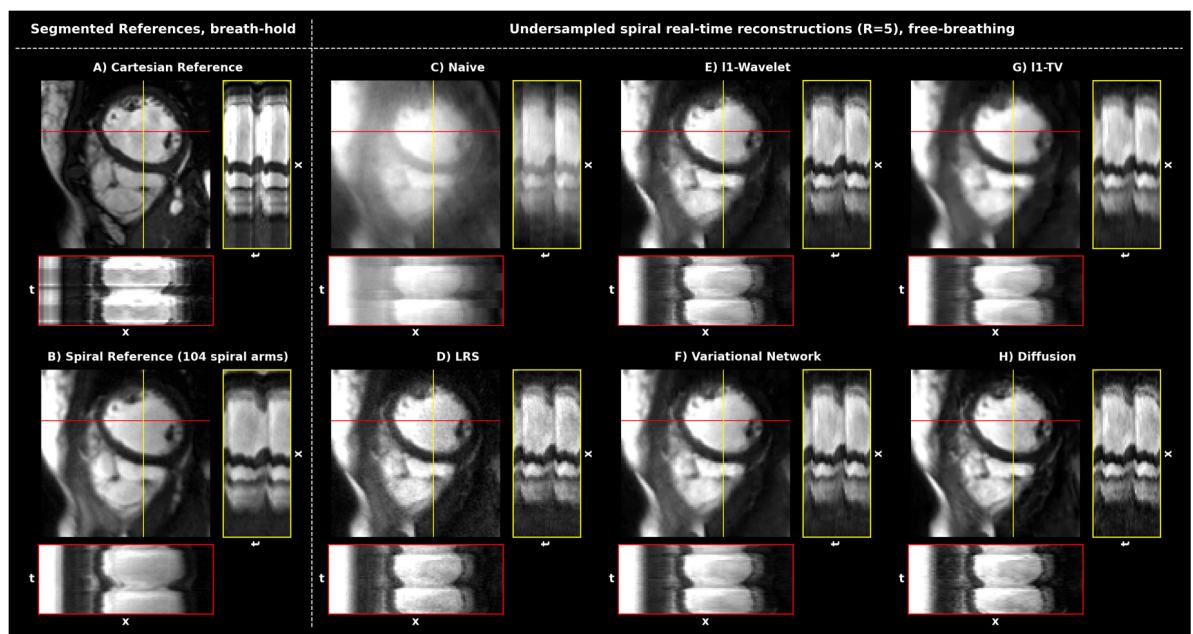

**Figure 6**: Spatio-temporal examination of the diastolic frames as presented in Fig. 4, centered on the myocardium. x-t plots represent 40 frames along the indicated vertical and horizontal lines. The Cartesian cine consisted of 21 frames, whereas the spiral cine contained 24 cardiac frames. The temporal evolution was partially repeated, respectively, to culminate in 40 frames. Visually, the diffusion reconstruction results in the sharpest depiction, especially also for the endocardial border.

Fig. 4 shows a comparison between different reconstruction approaches of a mid-ventricular slice in a healthy participant in systolic and diastolic view, respectively. The images on the left represent segmented / binned reconstruction (Cartesian and spiral) acquired in breath-hold. All remaining images show reconstructions of undersampled real-time frames in free-breathing, with the respective technique indicated. Spiral acquisitions were cropped to match the FOV of the Cartesian reference.

As typical for bSSFP-sequences, a strong contrast between the blood pool and the myocardium was obtained. Both segmented cine methods in Fig. 4 A) & B) provide a depiction free from apparent aliasing artefacts. For these acquisition at 1.5 T, bSSFP-signal voids could efficiently be minimized by shimming and therefore are mainly observable in the outer regions of the anatomy. The gold standard Cartesian method provides a slightly higher contrast and resolution than the spiral method which are caused by differences in the acquisition parameters, especially TE.

Increased sharpness of the reference might be due to off-resonance effects, which only lead to an image shift in Cartesian sampling, but spatial blurring for spiral trajectories[47]. Especially the vessels in the lung appear clearer for the standard technique.

In patients with arrhythmia (see Fig. 5), the segmented Cartesian method fails to correctly depict all cardiac phases due to temporal blurring induced by incorrect combination of data from different cardiac phases.

For real-time acquisitions in free-breathing, all reconstruction methods are essentially able to largely remove undersampling artefacts (which are apparent in the naïve image reconstruction, Fig. 4 C)). Differences between reconstructions methods can be seen more clearly in the cropped view depicted in Fig. 6. While the l1-TV method tends to spatially blur the image, the LRS method leads to an increase in noise. L1-WV, VN as well as the diffusion approach all offer high quality reconstructions, wherein both l1-WV and VN seem to be slightly more blurred than the diffusion model. Also, fine structures appear to be the sharpest for the latter technique. Small signal fluctuations due to blood flow can be observed in the x-t plots. At most, l1-WV results in slight "Wavelet-artefacts", apparent as blocky clusters of pixel.

Videos S1 and S2 in Supporting Information show a comparison between reconstructions of a sequence of 40 timeframes of one healthy participant and one arrhythmic patient in mid-ventricular SAX orientation. S3 and S4 in Supporting Information provide the corresponding results from diffusion-based reconstruction of the real-time acquisition in free-breathing for the entire stack from base to apex. In the healthy volunteer, cardiac motion can be accurately displayed by the Cartesian reference as well as the real-time reconstruction methods. In the patient, however, the segmented Cartesian acquisition fails and results in temporally blurred series, whereas the real-time acquisition precisely depict cardiac dynamics. Remaining artefacts from the pulsating blood flow are visible in all methods.

In the l1-WV reconstruction, again, wavelet artefacts are apparent to a minor degree. Higher noise in LRS covers artefacts but also fine details. Slight blurring impacts the VN reconstructions. Contrast and spatial sharpness seem to be best for the diffusion model. In the temporal view, the stochastic nature of the reconstruction, which was performed for each frame individually led to marginal noise-like artefacts in the time-series.

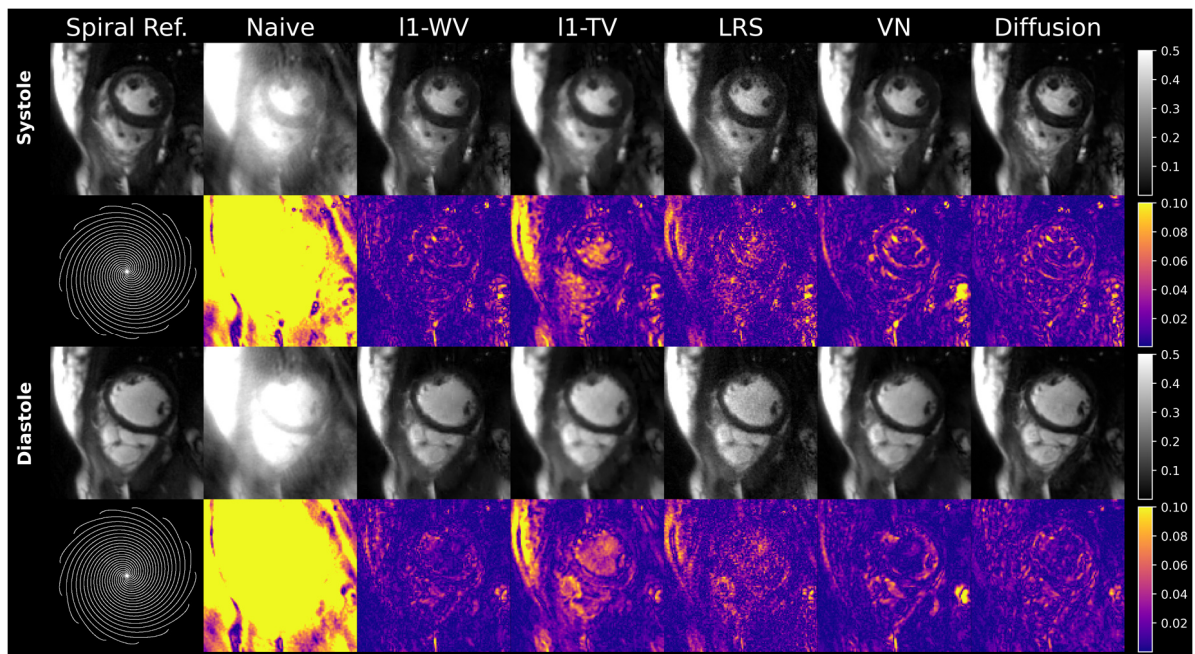

**Figure 7**: Simulation with retrospectively undersampled data. The undersampling pattern shown on the left side was applied to a fully sampled systolic and diastolic frame, which holds segmented data acquired in breath-hold across several heartbeats (i.e. Spiral Ref.). The images obtained from applying the different reconstruction techniques to the resulting undersampled k-spaces are depicted subsequently. Images were individually rescaled to a data range of [0 1]. The color encoded maps shows absolute difference images between the reference (Spiral Ref.) and corresponding reconstructions.

## 3.2 Quantitative comparison of reconstruction methods

For a quantitative evaluation, reconstructions of the proposed methods were generated following the description in Sec. 2.6. Images cropped to the region of the heart can be seen in Fig. 7 for a systolic and diastolic cine heart phase of one healthy volunteer. Corresponding quantitative metrics are listed in Table 2.

The difference images highlight remaining artefacts in the reconstructions of undersampled data. l1-TV differences appear more structural whereas LRS differences mainly occur noise-like. l1-WV, VN and diffusion images and difference images seem to be almost qualitatively equivalent and l1-WV and VN reconstruction appear slightly more blurred than the diffusion method. The latter finding is supported by the quantitative metrics. Even though the l1-wavelet reconstruction resulted in the best overall metrics (SSIM, NRMSE, PSNR), the metrics coincide within their respective standard deviation. Additionally, Tab. 2 lists the results of the randomized expert reader study (see Sec. 2.7). Overall differences in metrics between the real-

**Table 2: Results from the quantitative evaluation and the expert reader study**

|  | Cart. Cine | Naïve | L1-TV | L1-WV | LRS | VN | Diffusion |
|---|---|---|---|---|---|---|---|
| **Quantitative Metrics** | | | | | | | |
| **SSIM [%]** | - | 44.9±4.5 | 90.5±2.8 | 92.9±2.1 | 88.9±2.8 | 92.2±2.8 | 90.7±2.7 |
| **NRMSE [%]** | - | 107±25 | 14.6±6.8 | 8.6±2.0 | 12.1±3.5 | 9.4±3.3 | 9.3±1.8 |
| **PSNR [dB]** | - | 12.9±1.6 | 30.8±3.7 | 34.9±2.4 | 32.0±2.7 | 34.3±3.2 | 34.3±2.4 |
| **Expert Reader Study** | | | | | | | |
| **Myocardium-Blood contrast** | 1.0±0 | - | - | 1.3±0.5 | 1.3±0.5 | 1.1±0.3 | 1.1±0.3 |
| **Noise** | 1.0±0 | - | - | 1.9±0.5 | 2.5±0.5 | 1.2±0.4 | 1.8±0.6 |
| **Artefacts** | 1.4±0.7 | - | - | 2.5±0.9 | 1.6±0.7 | 1.8±0.6 | 2.0±0.6 |
| **Sharpness of myocardial contours** | 2.4±1.6 | - | - | 2.1±0.6 | 2.0±0.6 | 1.8±0.7 | 1.6±0.7 |
| **Dynamics** | 2.0±1.2 | - | - | 1.2±0.4 | 1.0±0 | 1.0±0 | 1.1±0.3 |
| **Sharpness of myocardial contours - Patients only** | 3.6±1.7 | | | 2.4±0.5 | 2.0±0.7 | 1.8±0.8 | 1.8±0.8 |
| **Dynamics - Patients only** | 3.0±1.2 | - | - | 1.0±0 | 1.0±0 | 1.0±0 | 1.0±0 |

Caption: Structural similarity index (SSIM), normalized root mean squared error (NRMSE) and peak signal-to-noise ratio (PSNR) with respective standard deviation were calculated from 1632 cine frames from 70 slices of six healthy participants. Using the data of 5 patients and 8 healthy participants, an expert reader study was performed on a five-point Likert-scale (1: best, 5: worst) for the Cartesian cine as well as, l1-WV, LRS, VN and diffusion real-time reconstructions.

time reconstructions are small, especially when considering the standard deviations. Myocardium-Blood contrast as well as the dynamics score show no dependency on the reconstruction method. The rating reflects that LRS reconstructions produced noisier images, whereas reconstructions using the VN show a better "Noise" score, which can be explained by its tendency to slightly blur the images. l1-Wavelet artefacts are on average more pronounced in comparison, with LRS depicting the best "Artefact" rating, probably due to artefacts being suppressed by the noise. Diffusion reconstructions yield on average slightly improved ratings in terms of "Sharpness of myocardial contours" in comparison to the other reconstruction methods.

Notable differences between the Cartesian gold-standard and the real-time free-breathing reconstructions occur. While the gold-standard produces images with less noise and artefacts, dynamic motion can be significantly corrupted. "Dynamics" as well as "Sharpness" ratings indicate that real-time reconstructions perform better than the Cartesian cine in depicting the cardiac movement especially when influenced by arrhythmia.

## 3.3 Cardiac functional parameters

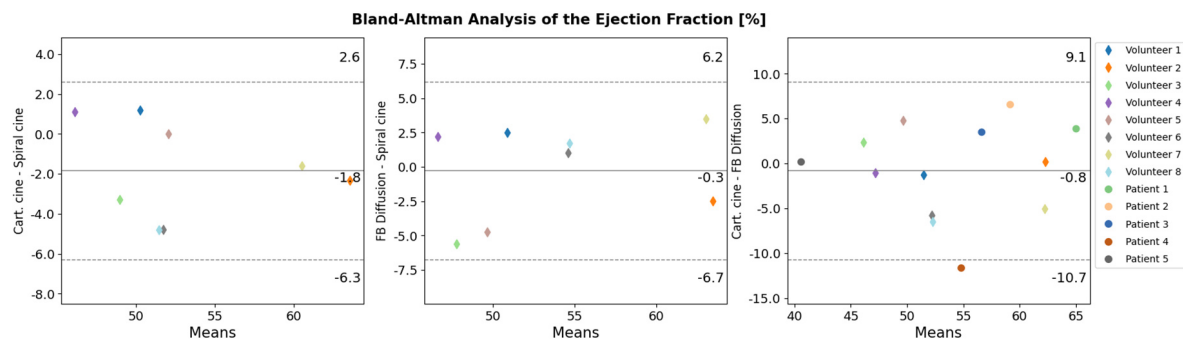

Figure 8: Bland-Altman Analysis comparing the ejection fraction between the two fully sampled methods acquired in breath-hold in a segmented fashion ("Cartesian cine" and "spiral cine") and respective results for the diffusion reconstruction of real-time, free-breathing acquisitions. The horizontal lines depict mean differences and ±1.96σ limits of agreement.

Fig. 8 shows the results from the Bland-Altman analysis depicting the results from volumetry according to Sec. 2.8. Ejection fraction show overall good agreement between the three methods, namely the Cartesian gold-standard (breath-hold, segmented acquisition), the self-gated spiral cine (breath-hold, segmented acquisition) and the spiral free-breathing real-time diffusion reconstruction. Differences in the absolute values of the EF are only small between volunteers and patients, so that the two groups do not separate horizontally. Ejection fraction evaluation of the two segmented methods, namely the Cartesian reference and the binned spiral cine, resulted in a small negative bias of -1.8±4.5%. Analogously comparing segmented spiral cine from breath-hold acquisitions with diffusion reconstruction of free-breathing acquisitions resulted in -0.3±6.5% mean differences of the EF. EF of Cartesian standard compared to free-breathing real-time depicts a bias of -0.8±9.9%. However, separating healthy volunteers results in a bias of -1.6±7.4%, while the isolated patient group depict a value of 0.5±13%. Slight mean deviations are observable for EDV, ESV and SV with 95% confidence intervals showing moderate uncertainties (as depicted in the full overview of the Bland-Altman Analysis shown in Supplementary Fig. 1).

### 3.4 Speed of acquisition & reconstruction

The acquisition time of the Cartesian gold standard method in breath-hold took between 4.0 s and 10 s per slice. Spiral acquisitions were performed for 4.5 s - 5.7 s in free-breathing and up to 15 s in breath hold. Keep in mind, that long acquisition times in breath hold for the spiral acquisitions were only needed for the generation of fully sampled data for training. Since real-time acquisitions with a temporal resolution of 48 ms are possible with the proposed sequence and corresponding reconstruction techniques, the total acquisition time could be significantly reduced by only acquiring data for roughly two heartbeats. Without ECG-gating and breath hold, the acquisition time, required for fully covering the whole heart would need less than one minute with the spiral sequence.

Nevertheless, reconstruction times are still an issue for the advanced reconstruction used in this study. GROG of a temporal average k-space took about 138 s and 4.3 s for a single real-time frame on CPU. Calculating coil sensitivity maps from the averaged k-space took about 3.8 s. The fastest runtime of the subsequently applied algorithms was possible with the LRS model, which needed 0.4 s per frame with an implementation on a GPU. VN also offered fast reconstructions with 0.9 s for passing off-grid data of one frame through the network. Since VN is the only method using GPU-based convolution gridding, the time consuming GROG didn't have to be taken into account here. The diffusion approach took about 25 s for the reconstruction of one real-time frame. Both l1-CS methods were restricted to run on CPU only, resulting in reconstruction times of 4.7 s for the l1-TV and 3.8 s for the l1-WV model. Moreover, the model-based CS-approaches do not need any time consuming training like the VN and the diffusion model.

## 4 Discussion

In this work, we aimed to investigate the potential of exploiting a diffusion probabilistic model for the reconstruction of undersampled "real-time" cardiac cMRI with spiral readouts. To this end, a score-based diffusion model was first trained with cardiac cine images and then integrated into a physics-based reconstruction scheme. The performance of this approach was compared to various state-of-the-art model-based and data-driven reconstruction techniques such as l1-compressed sensing (total variation and wavelet)[28], low rank plus sparse[27] and a Variational Network[16]. While the fully sampled Cartesian gold standard offered the highest image quality in the case of regular heartbeats, binning data of multiple heartbeats in patients suffering from arrhythmia, or, who cannot comply with repeated breath-hold acquisitions, is susceptible to allocation errors. In this regard, we showed that the proposed imaging technique facilitates an alternative approach to the clinical standard, offering comparable temporal and spatial resolution and, ultimately, allowing robust evaluation of cardiac functional parameters. But also for compliant patients with regular heartbeat, real-time cMRI in free-breathing can significantly increase patient comfort, reduce costs and indirectly make this high-quality examination accessible to a broader collective.

Diffusion probabilistic models are an excellent tool to generate new samples that lie within the probability distribution of a distinct data set. In our special use case, the score-based model[20,25] generates images, which are similar to the high-quality cine images used for training. By additionally enforcing fidelity to the (undersampled) measurement data and the coil sensitivities, accelerated scans can be transformed into images which are comparable to those obtained from fully sampled k-spaces. The stochastic and generative nature of the diffusion model was further investigated using repeated reconstructions of the same single image, where deviations in the "repetition domain" appeared as high-frequency noise. In the dynamic view of the real-time series, whose images were reconstructed individually by the diffusion model, very slight contrast changes can be seen at a few image positions across the temporal domain. Including temporal dependencies between adjacent frames into the diffusion prior, comparable to recently proposed video diffusion models[48], could help to provide an even more consistent reconstruction over time.

**Reconstruction speed**
One major disadvantage of current implementations of probabilistic diffusion models are certainly the demanding computational efforts and long inference times, as the reverse diffusion process is typically based on several hundreds of time steps. In our approach, we did not initialize the reconstruction with pure Gaussian noise, but with a preliminary reconstruction provided by iterative SENSE, to justify a significant reduction of the number of "iterations". Furthermore, GROG was used to move data to a Cartesian grid once, to avoid a gridding-bottleneck throughout the reverse time steps. Nevertheless, latencies of 25 s for a single frame are still too long to provide meaningful delays for clinical routine[49] and more conceptual approaches to speed up the reconstruction are needed.

Denoising Diffusion Implicit Models (DDIM) were proposed as a non-Markovian, deterministic alternative approach to traditional diffusion models. Adaptions of the DDIM framework for the reconstruction of undersampled MR-data were adapted in [50,51], achieving high image-qualities with reduced sampling steps and accelerated sampling times. Distillation techniques could also be applied in the context of diffusion models. Here the key idea is to train two separate models, namely a teacher model, which presents the complex iterative diffusion sampling process and student model, which aims to replicate the teacher model in a more efficient way, drastically reducing the number of required steps. An in-depth explanation on distillation of diffusion models can be found in [52-54]. Distillation techniques have also been proposed for MR reconstruction [55,56], but to our knowledge not yet in the context of diffusion models for undersampled MR reconstruction. Korkmaz et al.[57] proposed an unrolled approach to realize self-supervised MRI reconstruction using only 5 inference steps with a transformer architecture. The significant acceleration was here also made possible by initializing the reconstruction with a zero-filled estimate. In [24] the authors employed a large diffusion step size for rapid sampling. Since traditional diffusion models assume normality for the reverse transition, the authors employ an additional network to implicitly represent the distribution of a reverse diffusion step. Similarly, Zhao et al. [23] combine diffusion sampling and GAN's to achieve state-of-the-art reconstruction performance using only 16 diffusion steps.

While our work focused on developing "real-time" acquisitions that depict good image quality at a high temporal resolution for a dedicated point in time, authors of similar work partially restrict the term "real-time" to techniques that furthermore reduce addressed reconstruction latencies to a few hundred milliseconds after acquisition[49]. We also depicted reconstruction times for gpu-accelerated and solely cpu-based methods, which is definitely limited. We mainly intended to present an overview of reconstruction times during this study and also recognize that further optimizations may significantly reduce latencies.

**Evaluation of image quality and Bland-Altman-Analysis**

In our comparison to alternative methods, an advantage of the diffusion model was the ability to reconstruct somewhat sharper images, as indicated by the results of the expert reader study. Artefact level and SNR, however, were rated slightly better for the VN, such that no method showed overall clear superiority in their current state. In particular, when compared to the images obtained from corrupted segmented acquisition in case of arrhythmia, VN- and diffusion-based reconstructions of real-time data appear to be of comparable (high) quality.

The classical scalar metrics SSIM, NRMSE and PSNR are not really predestined to work out these subtle differences sensitively. Here, NRMSE and PSNR are more or less on par between VN and Diffusion, while SSIM is slightly lower for the latter. With no fully sampled ground truth data available for a single real-time frame, our calculations were based on a segmented spiral reference with data from multiple heartbeats. Slight variations in the heart rate may result in blurring and a loss of detail in the reference data. Consequently this leads to reduced metrics for methods that preserve details while improving metrics for ones that exhibit increased blurring. Interestingly, scalar image quality metrics were best for the l1-wavelet reconstruction, while performing either comparable or worse than the other reconstruction techniques in the expert reader study. With these quantitative metrics often solely being used for the evaluation of image quality, this discrepancy is an important issue.

Volumetric evaluation using Bland-Altman analysis showed on average agreement of Cartesian-gold standard and free-breathing real-time acquisition reconstructed using the diffusion model. Nevertheless significant uncertainties occurred, limiting conclusion that can be inferred from the values. For the segmented methods, such as the Cartesian standard and the segmented spiral acquisitions, only a singular heartbeat is available. Here averaging effects might influence the evaluation of the endocardial border in end-systolic cardiac phase, leading towards lower results with respect to the ejection fraction. Variability in the dynamic motion in different cardiac cycles could also affect the evaluation of the real-time approach. Even though multiple heartbeats were acquired only the endocardial border of a single end-diastolic/-systolic border was segmented by the examiner due to the extensive effort involved. However, segmenting multiple real-time heartbeats might also allow to evaluate the variations in the cardiac motion. Additionally, physical load due to breath-hold acquisitions versus free-breathing measurements might also influence volumetric parameters. A proper comparison would therefore necessitate a clinically validated real-time method, which was not accessible during the study period. Lastly, segmentations were performed by two examiners, introducing inter-variability as well as intra-variability in the evaluation.

We furthermore acknowledge that the evaluation dataset containing 8 healthy volunteers and 5 patients is rather small and a more rigorous analysis would profit from a larger subject cohort.

**Hyperparameters of diffusion models**

The choice of hyperparameters represent an important role when training and running a diffusion model for MRI reconstruction. Noise scheduler and number of perturbation steps define the main properties of diffusion process during training. During inference we guided the diffusion model with heuristically chosen weights for the data consistency terms inspired by Jalal et al.[19]. Emerging theoretical descriptions depict an implementation of the generative diffusion prior for the solution of inverse problems in a fashion more in line with the mathematical background[40,41]. In the future, these might help to overcome a heuristically chosen combination of data generation prior and data consistency conditioning.

An accelerated inference was achieved by initializing the diffusion model with a coarse SENSE reconstruction and reduction of the number of diffusion steps, similarly to Chung et al.[22]. Noise scheduler, number of perturbation steps, data consistency weighting, initialization and number of diffusion steps during inference all have severe influence on the final reconstruction quality. Moreover, diffusion models generally utilize an l2-loss function, based on their mathematical framework. However, the choice of loss function in other data driven models presents an additional parameter. For example MSE-loss could have induced additional blurring in the VN images. Nevertheless Kleineisel et al.[16] showed that differences in trainings of VNs using SSIM-loss versus MSE-loss were only small.

Rigorous optimization of all these parameters as well as exploration of various network architectures might further yield improved image qualities, but also require extensive effort which was beyond the scope of the current study.

Differences between the data-driven methods, namely the VN and the diffusion model were low, which does not currently justify the great effort involved in training and the comparatively long computing times of the latter.

Nevertheless, due to the very promising and versatile properties also shown in this paper, diffusion models in medical imaging are definitely on the rise [18] and continuous improvements, regarding inference speed and image quality are under intensive investigation[58–61].

Our code will be available upon publication at https://github.com/expRad/Diffusion.

## 5  Conclusion

We proposed a diffusion probabilistic model to reconstruct undersampled real-time cardiac cine MRI based on spiral trajectories acquired in free-breathing as a promising strategy to provide shorter and more robust cMRI exams, especially in patients with irregular cardiac cycles and who cannot hold breath properly. In a comparison with alternative state-of-the-art reconstruction techniques, the diffusion model showed potential to improve sharpness, however, differences were rather small and come at the cost of extensive training effort as well as the long inference times of the current implementation.

**Availability of Supporting Videos S1-S4**
For access to referenced video files, feel free to contact us via schad_o@ukw.de.

**Acknowledgement**

We thank Chung et al. https://github.com/HJ-harry/score-MRI and Song et al. https://github.com/yang-song/score_sde for providing a code base for score-based generative diffusion models.

# Supporting Information S1

**Full Overview of the Bland-Altman Analysis**

Supplementary Figure 1 presents a full overview of the volumetric evaluation described in section 2.8 in the main text, additionally including analysis of the end-diastolic, end-systolic and stroke volume.

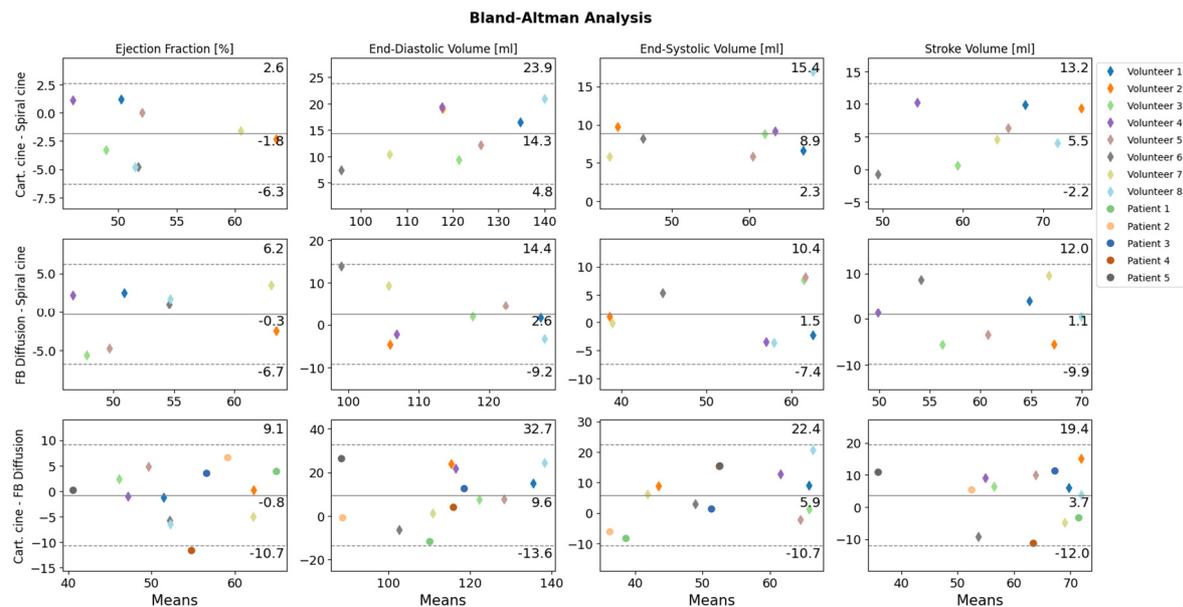

Supplementary Figure 1: Full overview of the volumetric evaluation using Bland-Altman Analysis comparing the two segmented methods acquired in breath-hold, namely the Cartesian ECG-gated reference and the self-gated spiral cine and respective results for the diffusion reconstruction of real-time, free-breathing acquisitions. The horizontal lines depict mean differences and ±1.96σ limits of agreement.